\documentclass{elsart4-1}

\usepackage{amsmath, amssymb}

\usepackage[english,highlight]{LLNdef}
\usepackage{journames}
\usepackage[highlight]{modifications}
\usepackage{cite}
\usepackage{citesort}

\ifx\pdfoutput\undefined\usepackage{graphicx}\else\usepackage{graphicx}\fi
\graphicspath{{img/150dpi/}}

\usepackage[english,francais]{babel}

\newtheorem{e-proposition}[theorem]{Proposition}

\newtheorem{e-definition}[theorem]{Definition\rm}

\def\og{\leavevmode\raise.3ex\hbox{$\scriptscriptstyle\langle\!\langle$~}}
\def\fg{\leavevmode\raise.3ex\hbox{~$\!\scriptscriptstyle\,\rangle\!\rangle$}}

\begin{document}

\begin{frontmatter}


\selectlanguage{english}
\title{Magnetism in reduced dimensions}

\vspace{-2.6cm}

\selectlanguage{francais}
\title{Magn{\'e}tisme en dimensions r{\'e}duites}


\selectlanguage{english}
\author[of]{Olivier Fruchart}
\ead{Olivier.Fruchart@grenoble.cnrs.fr}
\author[at]{Andr{\'e} Thiaville} \ead{thiaville@lps.u-psud.fr}

\address[of]{Laboratoire Louis N{\'e}el (CNRS) -- 25, avenue des Martyrs -- BP166 -- F-38042 Grenoble Cedex 9 -- France}
\address[at]{Laboratoire de Physique des Solides -- Universit{\'e} Paris-Sud, B{\^a}t.510 -- 91405 Orsay Cedex -- France}

\begin{abstract}
We propose a short overview of a few selected issues of magnetism in reduced dimensions, which are
the most relevant to set the background for more specialized contributions to the present Special
Issue. Magnetic anisotropy in reduced dimensions is discussed, on a theoretical basis, then with
experimental reports and views from surface to single-atom anisotropy. Then conventional
magnetization states are reviewed, including macrospins, single domains, multidomains, and domain
walls in stripes. Dipolar coupling is examined for lateral interactions in arrays, and for
interlayer interactions in films and dots. Finally thermally-assisted magnetization reversal and
superparamagnetism are presented. For each topic we sought a balance between well established
knowledge and recent developments.

{\it To cite this article:  O. Fruchart, A. Thiaville, C. R. Physique X (y) (2005).}

\vskip 0.5\baselineskip

\selectlanguage{francais}
\noindent{\bf R{\'e}sum{\'e}}%
\vskip 0.5\baselineskip%
\noindent%
Nous proposons un panorama de quelques aspects du magn{\'e}tisme en dimensions r{\'e}duites, appropri{\'e}s
comme toile de fond pour les articles plus sp{\'e}cialis{\'e}s de ce num{\'e}ro sp{\'e}cial. L'anisotropie
magn{\'e}tique en dimensions r{\'e}duites est discut{\'e}e, sur le plan th{\'e}orique, puis appuy{\'e}e par des
exemples, allant des surfaces aux atomes individuels. Les configurations d'aimantation les plus
courantes sont ensuite d{\'e}crites: macrospins, monodomaines, multidomaines, parois dans des bandes.
Les couplages magn{\'e}tiques, essentiellement dipolaires, sont d{\'e}crit pour des r{\'e}seaux et pour des
bi-couches. Enfin nous pr{\'e}sentons les effets de l'activation thermique, de la baisse de
coercitivit{\'e} jusqu'au superparamagn{\'e}tisme. Pour chaque aspect nous avons recherch{\'e} un {\'e}quilibre
entre r{\'e}sultats {\'e}tablis et d{\'e}veloppements r{\'e}cents.

{\it Pour citer cet article~: O. Fruchart, A. Thiaville, C. R. Physique X (y) (2005).}

\vskip 0.5\baselineskip%
\noindent{\small{\it Mots-cl{\'e}s~: Nanomagn{\'e}tisme, Micromagn{\'e}tisme, Anisotropie magn{\'e}tique, Superparamagn{\'e}tisme, Dimension r{\'e}duite} }%
}\end{abstract}
\end{frontmatter}

\def\Olivier{{\color{blue} \texttt{[Olivier]}}}%
\def\Andre{{\color{red} \texttt{[Andr{\'e}]}}}%
\def\OlivierAndre{[{\color{blue} \texttt{Olivier}}/{\color{red} \texttt{[Andr{\'e}]}}]}%
\def\MAE{MAE\/\xspace}

\selectlanguage{english}
\section{Introduction}

Magnetism in reduced dimensions has been an active topic in the last two decades. Much progress,
still under way, has been made possible by the conjunction of three aspects. First, the progress
of fabrication techniques, both deposition and lithography. Second, the progress of magnetic
characterization techniques like XMCD and XMLD, Lorentz microscopy, SEMPA, XMCD/XMLD-PEEM, SPLEEM,
sp-STM, magnetic scattering and surface diffraction etc. Third, the considerable increase of
computing power. Today all three aspects overlap in the range $\unit[20]{nm}-\thickmicron{1}$,
which makes our era very productive. This length scale could define \emph{nanomagnetism}. A better
term might have been \emph{mesomagnetism}, \ie at the cross-over from macroscopic behaviors to
uniform magnetization, although the term 'meso' has not been considered by the community of
magnetism.

The interest in nanomagnetism has also been boosted by the discovery of new (or revisiting of)
phenomena that arise owing to the fabrication of heterostructures at the nanoscale, and that
underlie most of the topics of the Special Issue: giant magnetoresistance, tunneling
magnetoresistance, exchange anisotropy and bias, spin torque. In this contribution we review some
basic aspects of magnetism in reduced dimensions for mostly single systems, that are useful to
consider before implementing some of the above-mentioned effects in complex heterostructures, may
it be for realizing functional devices or structures for fundamental investigations. The topics
covered are magnetic anisotropy, magnetization states, interactions (mostly dipolar) and thermal
activation.

\section{Magnetic anisotropy in low dimensions}

Here we discuss \textsl{microscopic} magnetic anisotropy energy~(\MAE), a field subject to
breaking discoveries in the recent years. Dipolar anisotropy will be treated in
\secref{sec-MagnetizationStates}. Other low-dimensional effects are excluded from the discussion,
such as magnetic moments at interfaces or the reduction of ordering temperature. See
\cite{bib-GRA93,bib-SIE92,bib-POU99} for reviews. The former is relevant to spintronics \eg for
the TMR effects, see Ref.\cite{bib-SCH05}. The latter is often screened by
\emph{superparamagnetism}, treated in \secref{subsec-Superparamagnetism}.

\subsection{Theoretical descriptions}

MAE results from the interaction of magnetization with the local environnement of atoms, via the
crystal electric field\cite{bib-SAN04}. In bulk materials at equilibrium this is the
\textsl{magnetocrystalline anisotropy}~$\Emc$. When any dimension of a system is reduced
corrections to $\Emc$ arise, due to interface or strain~(deformation of the structure).

Long before thin films could be grown epitaxially and at the nanometer scale, N{\'e}el foresaw that
the local breaking of symmetry at surfaces should induce a correction to $\Emc$, which he named
\emph{surface anisotropy}~($\Es$, an energy per unit area)\cite{bib-NEE54}. $\Es$ is nowadays
often referred to as \emph{N{\'e}el surface anisotropy}\cite{bib-dipolarSurfaceAnisotropy}. He used a
pair model to predict the angular variation of $\Es$, the summation being restricted to magnetic
neighbors. $\Es$ could be expanded in spherical harmonics, although simple polynomial expansions
are more popular, with the most simple form being uniaxial anisotropy: $\Es=\Ks \cos^2\theta$. In
a crude approximation numerical coefficients were derived from magnetoelastic coupling
coefficients, yielding values around $\unit[0.1-1]{mJ/m^2}\sim\unit[0.1-1]{meV/atom}$,
surprisingly of the correct experimental order of magnitude, $\unit[0.1]{mJ/m^2}$ as revealed
experimentally much later. It is commonly acknowledged today that this model fails to predict
exact figures, even their sign, which can only be derived from experiments or \abinitio
calculations.

A more rigorous view of surface anisotropy than N{\'e}el's was given by Bruno, who predicted the
proportionality of surface anisotropy constants with the anisotropy of the angular
momentum\cite{bib-BRU89b}. To understand this fact, it should be recalled first that in a bulk 3d
solid the orbital momentum is very nearly zero, as the electron wavefunctions loose the rotation
invariance that exists in the atom, because of the crystalline electric field. As a result, the
angular momentum in bulk 3d is very small compared to the spin moment, and in fact appears only as
a perturbation when including the spin-orbit term. At a surface or interface however, the
crystalline electric field looses symmetry and becomes compatible with a perpendicular orbital
moment. This induces, via the spin-orbit coupling, an extra MAE. The initial model of
Bruno\cite{bib-BRU89b} was based on a tight binding approach of the electronic structure in a $3d$
transition metal atomic layer~(AL), and has been refined later\cite{vanderLaan98}. More realistic
\abinitio calculations have revealed some departures from this general trend\cite{Ravindran01}.

\subsection{Thin films, a model for surface anisotropy}

Thin epitaxial films are model systems. The translation symmetry in-the-plane yields a
laterally-small unit cell, at reach to \abinitio computation, and easing experimental analysis.
Close-to-ideal films are nowadays routinely fabricated for many systems, which can be controlled
down to the single \AL.

The first clear confirmation of the existence of $\Es$ was given by Gradmann \etal in the late
sixties\cite{bib-GRA68}. The total uniaxial \MAE $E=K_\mathrm{tot}(t)\cos^2\theta$ of
$\mathrm{Fe}_{52}\mathrm{Ni}_{48}/\mathrm{Cu}(111)$ films of thickness $t$ followed a $1/t$
dependence, the slope being ascribed to $\Es$: $K_\mathrm{tot} t=K_\mathrm{bulk} t+2\Ks$. Notice
that $K_\mathrm{bulk}$ includes both magnetocrystalline anisotropy and shape anisotropy
$\Kd=\frac12\muZero\Ms^2$, $\theta$ being the angle of magnetization with the normal to the film
plane. These experiments, performed down to a few ALs, first suggested the possibility to attain
an effective \emph{perpendicular anisotropy}, provided that $\Ks$ is negative and sufficiently
large to overcome $\Kd$ for a few atomic layers. Examples of perpendicular anisotropy are
Au/Co/Au, Pt/Co/Pt and Pd/Co/Pd (multi)layers, with critical thicknesses for spin reorientation
transition in the range \thicknm{1-2}.

It was then realized\cite{bib-CHA88} that $1/t$ plots mix surface $\Ks$ and magnetoelastic $\Kmel$
contributions. Indeed structural models predict at equilibrium a $1/t$ relaxation of strain
$\overline{\overline{\epsilon}}$ in heteroepitaxial growth\cite{bib-JES67,bib-GRA74} so that
$\Kmel\sim1/t$. Thus true $\Ks$ values could only be extracted after substraction of $\Kmel$ when
$\overline{\overline{\epsilon}}$ is measured and magnetoelastic coefficients $\Bmel$ are known, or
in the more rare case of pseudomorphic growth over a range of many ALs, like for
Ni/Cu(001)\cite{bib-JUN94}. $\Ks$ values obtained in this fashion are reviewed in
Ref.\cite{bib-GRA93}.

More recently the direct measurement of $\Emel$ in films using bending cantilevers, and the
revisiting of previous data, revealed that $\Emel$ is no more linear with
$\overline{\overline{\epsilon}}$ in ultrathin films\cite{bib-SAN99,bib-SAN02b,bib-HA99,bib-KOM00}.
Higher order terms in $\epsilon$ need to be considered, which \eg for Fe can reverse the sign of
$\Bmel$ at less than $\unit[1]{\%}$ of strain\cite{bib-SAN02b}. This had been overlooked in bulk
samples because plastic deformation occurs well below the strain values commonly observed in
heteroepitaxial films. The reentrant in-plane magnetization of Ni/Cu(001) in the ultrathin
range\cite{bib-HA99,bib-GUT00}, is now explained by non-linear magnetoelastic effects. The strain
dependance of $\Ks$ itself was also postulated, initially on Ni/Cu(001)\cite{bib-BOC96}, however
of puzzlingly high magnitude, and could never be confirmed unambiguously. From all this it must be
concluded that magnetoelastic and true N{\'e}el-type anisotropy are entangled in thin films. Their
clear separation, even conceptually, is impossible in most systems, where only an effective $\Ks$
can be deduced from $1/t$ plots.

On the microscopic level several experiments~(see \cite{bib-WEL95} for the pioneering work) have
confirmed the link between MAE and the anisotropy of the orbital momentum, using magnetic circular
dichroism effects with soft X rays~(XMCD). The anisotropy of the orbital momentum for $3d$
elements at surfaces is of the order of $\unit[0.1]{\muB/atom}$.

\subsection{Surface anisotropy in nanostructures}

Beyond the model case of thin films, surface anisotropy applies to all atoms located at the
surface of any nanostructure. The length scales of the physical effects giving rise to $\Es$ are
in the low nanometer range. Thus the atomic arrangement close to the interface is crucial, so that
nanostructures fabricated by lithography or by any other artificial mean are not adequate to
evidence~$\Es$ in reduced lateral dimensions. Instead, when this field has been explored in the
last decade, one used \eg clusters fabricated by physical means\cite{bib-BAN05}, or epitaxial
self-organization~(\SO) at surfaces\cite{bib-FRU05b,bib-FRU05c}. The disentanglement of
magnetoelastic and true N{\'e}el anisotropy is even more difficult than for thin films, given the
complexity of geometry and strain, and \emph{in most cases} because of the distribution of local
environments~(loss of the small unit cell). Therefore, in the following we should consider $\Es$
as an \emph{effective} surface anisotropy, without trying to discuss its physical origin.

\begin{figure}[b]
  \begin{center}
  \includegraphics[width=153mm]{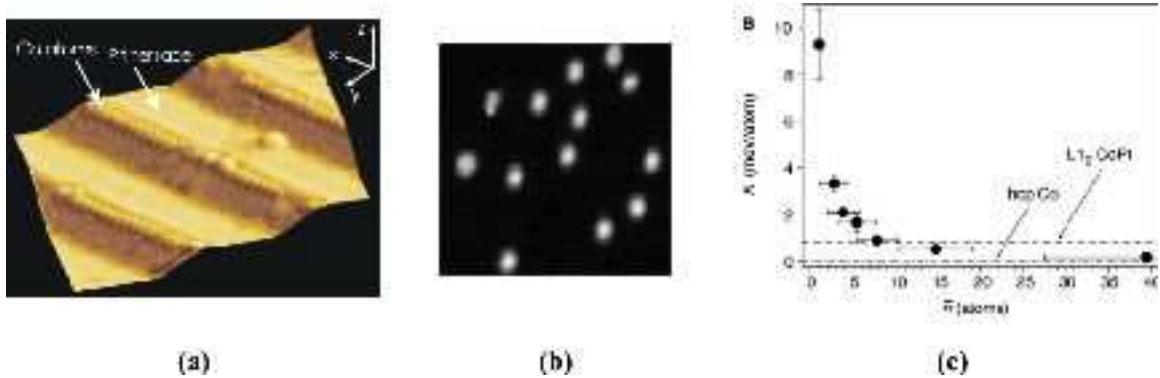}
  \end{center}
  \caption{\label{fig-SO-Co}(a)~Monoatomic Co wires decorating steps of Pt(997)\cite{bib-GAM02}
    (b)~Single Co atoms on Pt(111)\cite{bib-GAM03} (c)~perpendicular MAE for Co/Pt(111)
    as a function of the cluster size.}
\end{figure}

Notice that in nanostructures like those discussed above, the local reduction of dimensionality
can be more severe than at the 2D surface of thin films, \ie with a higher loss of coordination.
Epitaxial growth was then used for its ability to produce nanostructures with a more monodisperse
type of interfacial atoms than for clusters, to analyze quantitatively the concepts of \emph{edge
anisotropy} for a 1D interface~(\eg an atomic edge, or the edge of a monolayer-high island), or
even \emph{kink} anisotropy for a 0D defect along such a 1D interface, or an isolated magnetic
atom on a surface, as we will see. Pioneering work was performed on ultrathin films grown on
\emph{vicinal} surfaces, giving rise to a regular array of stepped
sites\cite{bib-ALB92,bib-WEB96}. After correction for the tilt of crystal axes for $\Emc$, a clear
linear variation of anisotropy with the miscut angle can be evidenced, and interpreted as a
\emph{step} anisotropy with a magnitude of the order of $\unit[1]{mJ/m^2}$. Later \SO
nanostructures have been used to further decrease the dimensionality, that were mainly studied
with X-ray magnetic circular dichroism~(XMCD), for its sensitivity and ability to yield the
orbital momentum and its anisotropy. Upon sub-\AL deposition on the vicinal surface Pt(997),
\thickAL{1}-high Co stripes of adjustable width were fabricated by step
decoration\cite{bib-DAL00}\bracketsubfigref{fig-SO-Co}{a}. A surface-RKKY-type of variation of the
\MAE was evidenced, oscillating with the width of the stripes\cite{bib-GAM04,bib-GAM03b} and
culminating for monoatomic wires to $\unit[2]{meV/atom}$\cite{bib-GAM02}. Minute amounts of Co
were also deposited around \tempK{15} on Pt(111), remaining as individual atoms because surface
diffusion is frozen at this temperature\bracketsubfigref{fig-SO-Co}{b}. A giant \MAE of
$\unit[9]{meV/atom}$ was measured. Upon annealing Ostwald ripening sets in, yielding islands of
well-controlled size and narrow size distribution. Thus, for Co in contact with Pt the variation
of \MAE from single atoms to bulk was fully spanned for the first
time\cite{bib-GAM03,bib-GAM05}~(\subfigref{fig-SO-Co}c)\cite{bib-XMCD-on-Dots}. These studies
confirm that the magnitude of $\Es$ increases dramatically from surfaces, to steps, then to kinks
or atoms. Besides, while the \MAE was derived directly from the fit of XMCD hysteresis curves, the
orbital moment was also measured, showing a similar increase for decreasing dimensionality. A
reasonable linear variation of \MAE with the \emph{anisotropy} of the orbital momentum is found
following the simple arguments from Bruno\cite{bib-BRU89b}. Ab initio calculations of clusters
have also shown this trend\cite{Guirado03}. Finally, notice the sharp decrease as a function of
size concerning orbital momentum and MAE: a bi-atomic island behaves closer to an infinite
monoatomic-wide wire than to a single atom, and bi-atomic wires are closer to a monolayer film
than to a mono-atomic wire\brackettabref{tab-coOrbitalMoment}. For 3D clusters ($\approx 3$~nm)
elaborated in the gas phase and measured individually (see Sec.~\ref{sec-macrospin}), the careful
analysis of the measured MAE has shown that surface terms also dominate\cite{Jamet01}.

{
\begin{table}[t]
  \centering
  \caption{Orbital momentum and magnetic anisotropy energy~(MAE) of Co atoms on Pt as a function of coordination
    (after \cite{bib-GAM02,bib-GAM03}).}
  \label{tab-coOrbitalMoment}
  \newlength{\widthItem}
  \def\textItem{~~Orbital momentum~~}
  \settowidth{\widthItem}{\textItem}
  \newlength{\widthItemm}
  \def\textItem{~~bi-atomic~~}
  \settowidth{\widthItemm}{\textItem}
  \newlength{\widthItemmm}
  \def\textItem{~~mono-atomic~~}
  \settowidth{\widthItemmm}{\textItem}
  \begin{tabular}{p{\widthItem}ccp{\widthItemm}p{\widthItemmm}cc}
    \hline\hline    &  ~~bulk~~  & ~~mono-layer~~ & \centering ~~bi-atomic wire~~  & \centering ~~mono-atomic wire~~  & ~~two atoms~~ &
    ~~single atom~~\\
    \hline Orbital momentum ($\muB/\mathrm{at}$) & 0.14 & 0.31 & \centering 0.37 & \centering 0.68 & 0.78 & 1.13 \\
    MAE ($\mathrm{meV}/\mathrm{at}$) & 0.04 & 0.14 & \centering 0.34 & \centering 2.0 & 3.4 & 9.2
    \\\hline
  \end{tabular}
\end{table}
}

\section{Magnetization states and magnetization processes in single systems}
\label{sec-MagnetizationStates}

\subsection{Basics of micromagnetism}
\label{sec-BasicsMicromagnetism}

A general introduction to the micromagnetic theory should be sought elsewhere\cite{bib-HUB98b}.
Here we discuss a few selected issues only.

\begin{figure}[b]
  \begin{center}
  \includegraphics[width=160mm]{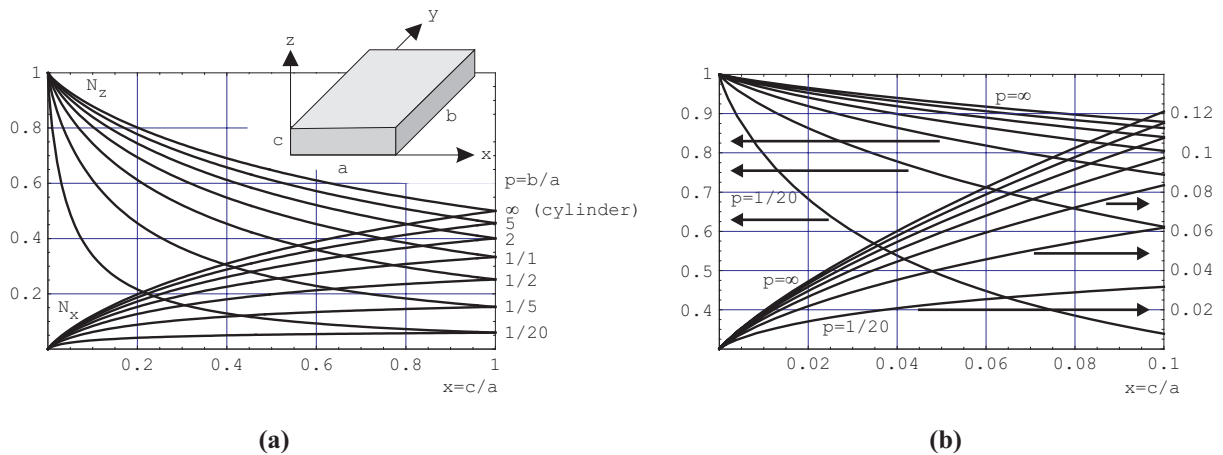}%
  \end{center}
  \caption{\label{fig-demag}Demagnetizing coefficients of prisms: (a)~arbitrary shape
    (b)~close-up view for flat prisms}
\end{figure}

Demagnetizing coefficients and magnetic length scales are useful parameters to discuss
magnetization patterns. It can be shown\cite{bib-BEL03} that a demagnetizing tensor
$\overline{\overline{\vect{N}}}$ can be defined for a sample of \emph{arbitrary} shape
\emph{assumed} to be uniformly magnetized:

\begin{equation}
  \label{eq-DemagCoef}
  <\vect{H}_\mathrm{d}(\vect{r})>=-\overline{\overline{\vect{N}}}.\vect{M}
\end{equation}

with $\vect{M}$ the magnetization vector and $<\vect{H}_\mathrm{d}(\vect{r})>$ the demagnetizing
field \emph{averaged} over the sample. The density of demagnetizing energy is immediately
$\Ed=-\frac12\muZero<\vect{H}_\mathrm{d}(\vect{r})>\vect M$. $\overline{\overline{\vect{N}}}$ is
positive and symmetric, thus can be diagonalized, so that along any main axis~$i$, one has
$<\vect{H}_\mathrm{d}(\vect{r})>=-N_i\vect{M}$. It can be shown that
$\mathrm{Tr}\overline{\overline{\vect{N}}}=1$, so that $\sum_{i=1}^3 N_i=1$. The emphasis is often
put on samples bounded by surfaces of polynomial equations not greater than two~(of practical
interest are thin films--also called slabs, ellipsoids, infinite cylinders with elliptical
cross-section). Only in these is $\vect{H}_\mathrm{d}$ uniform if $\vect{M}(\vect{r})\equiv
\vect{M}$, so that \eqnref{eq-DemagCoef} is valid at any point and the uniformity of
$\vect{M}(\vect{r})$ can be practically achieved along the main axes for $|\Hext|\gtrsim N_i M$.
Analytical formulas for $N_i$'s can be found for revolution ellipsoids\cite{bib-STO45},
prisms\cite{bib-RHO54,bib-AHA98}\bracketfigref{fig-demag}, cylinders of finite
length\cite{bib-ROW56,bib-RHO56,bib-GOO03}, and tetrahedrons\cite{bib-ROW92,bib-BEL03}. For other
geometries micromagnetic codes or Fourier-space computations\cite{bib-BEL03} can be used.

Characteristic magnetic length scales arise in non-homogeneous magnetization structures resulting
from the competition between two (or more) types of energy. The competition of exchange $A$ and
anisotropy $K$ yields the so-called Bloch wall width $\Delta=\sqrt{A/K}$ for the case of uniaxial
anisotropy. $\Delta$ is relevant to describe the width of walls when $\Ed$ is negligible, \eg in
the bulk or in ultrathin films of high anisotropy. The various definitions of the wall width are
reviewed in \cite{bib-HUB98b}, p.219. The competition of exchange and dipolar energy yields the
so-called exchange length $\Lambda=\sqrt{A/K_\mathrm{d}}$ with $\Kd=\frac12\muZero\Ms^2$.
$\Lambda$ is for instance a measure of the diameter of the core of magnetic vortices in
flux-closure patterns. One also defines the dimensionless \emph{quality factor} $Q=K/\Kd$. This
films of materials with perpendicular magnetocrystalline anisotropy support fully perpendicular
domains under zero external fields for $Q > 1$ (stripes and bubbles for low coercivity, up to
fully remanent for coercive materials), and continuously rotating structures for $Q < 1$ (weak,
strong stripes).

\subsection{Macrospins}
\label{sec-macrospin}

In particles of extremely small size the exchange energy dominates over all other energy terms so
that the magnetization state is always nearly uniform even during magnetization reversal, which
proceeds by \emph{coherent rotation} of all magnetic moments. This occurs for dimensions of the
order or below $\Lambda$, $\simeq\thicknm{10}$ for common materials like 3d magnetic metals. The
particle can then be reasonably described by a single magnetic moment, the so-called
\emph{macrospin}, subjected to an effective MAE that gathers the contributions from crystalline,
surface and shape anisotropies. The seminal paper investigating the magnetization reversal of
macrospins\cite{bib-STO48}, still used intensely, predicts that magnetization reverses by
reversible rotation and irreversible jumps, the latter occurring at field values that depend
strongly on the field angle with respect to that (those) of the effective anisotropy. This model,
initially developed for a uniaxial anisotropy of degree 2, was recently generalized to arbitrary
anisotropy\cite{bib-THI00}.

Experiments on individual nanoparticles of decreasing size, mainly performed by W. Wernsdorfer
with a technique called micro-SQUID\cite{bib-WER01}, have beautifully shown this behavior. The
anisotropy was revealed by the surfaces (in the space of the applied fields) where a jump occurs,
known generally as astroid\cite{bib-SLO56}. The measurements have been extended to dynamics. In
the slow regime dominated by thermal agitation, the magnetic relaxation was shown to involve only
one time constant at small sizes, whereas at larger sizes a non-exponential relaxation had been
observed \cite{Lederman94}. The former corresponds to the thermodynamic model of a particle in the
macrospin approximation, called N{\'e}el-Brown model\cite{bib-BRO63}, see
\secref{subsec-ThermalActivation}. In the fastest regimes in which magnetization precession is
important, \ie $t\lesssim\unit[1]{ns}$, it was shown that the application of pulses of
radio-frequency~(rf) fields could decrease the switching field of the particle if of the right
frequency \cite{Thirion03}, pointing to a non-linear resonance effect, i.e. the precession of the
macrospin driven by the rf field.

\subsection{Single domain states}

In-between the macrospin state and a macroscopic state where magnetic domains separated by domain
walls occur, lies the so-called \emph{single-domain state}. A single-domain state may be defined
by a state close to uniformly magnetized 'on the average', \ie displaying no magnetic wall not
vortex. The multidomain-to-single-domain transition was predicted long ago\cite{Kittel46} through
a comparison of the magnetostatic energy of a uniformly magnetized particle (proportional to its
volume) to the wall energy of a multidomain state (proportional to the particle surface). The
distinction between macrospin and single domain was introduced early, and some analytical
estimates of both sizes obtained\cite{Aharoni96}. For finite size and definite shapes, a number of
`phase diagrams' have been computed, that predict the magnetic structure of minimum energy as a
function of sample dimensions or anisotropy, for cubes\cite{bib-SCH88,bib-RAV98b},
disks\cite{Cowburn99}, squares\cite{bib-COW98}, rectangles\cite{bib-RAV00}. They all show that the
single domain state is reached at small sizes and large anisotropy, as expected.

\begin{figure}
  \begin{center}
  \includegraphics[width=110mm]{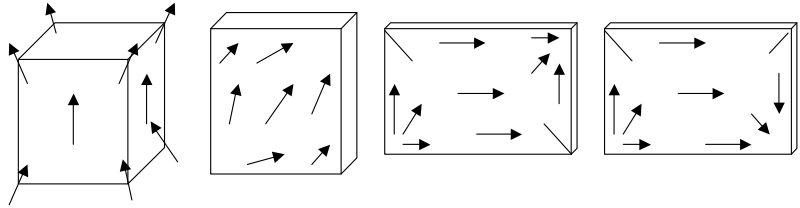}
  \end{center}
  \caption{\label{fig-states}The most common single domain states (schematic).
  From left to right: flower, leaf, S and C states}
\end{figure}

The term `single domain state' should not be confused with `uniform magnetization'. In single
domain states, although walls and vortices are not found at equilibrium, they may occur during
magnetization reversal through complex nucleation-propagation mechanisms. The critical sizes for
single-domain and macrospin are comparable for 3D compact particles, however the former may by far
exceed $\Lambda$ for high aspect ratios, like for thin flat dots. These dots correspond to the
majority of the small magnetic samples produced by the `top-down' approach, hence their detailed
study in the recent years. We mentioned in \secref{sec-BasicsMicromagnetism} that only for samples
shaped as surfaces of degree $\leq2$ is $\vect{H}_\mathrm{d}(\vect{r})$ uniform when the
magnetization distribution is so. For any other shape $\vect{H}_\mathrm{d}(r)$ is in general not
uniform, so that strictly speaking uniform magnetization cannot be achieved for whatever high
value of the applied field. The non-uniformity is especially strong for the case of thin and flat
elements~(including with in-plane elliptical shape), with a local magnitude that can be
considerably higher than the average value, especially close to the edges.  The deviations remain
down to infinitely small samples\cite{Aharoni91}, where they scale as $(\mathrm{size} /
\Lambda)^2$ \cite{Thiaville98}. The deformations of magnetization linked with the sample shape
have been transcribed in the names given to the configurations~\bracketfigref{fig-states}.

These deformations are very important as they control the orientation of the average magnetization
for magnetically-soft materials, through what has been called the configurational
anisotropy\cite{bib-COW98c}. This energy describes the tendency for the magnetization to become
non uniform within the sample so as to decrease magnetostatic energy at the minimum cost in
exchange energy. The rigorous computation of this energy requires a special micromagnetics
technique called path method \cite{Dittrich03}. Spectacularly enough, this energy explains how
apparent anisotropies of high degree can develop and be measured in triangles, pentagons etc.
whereas the conventional shape MAE is only of second degree in magnetization\cite{bib-COW00}. The
magnetization non-uniformity affects also greatly the magnetization reversal. Indeed, the
deviations are amplified when a field antiparallel to the average moment is applied. This results
in increased switching field and time for switching \cite{Nakatani89,Yang96} as well as non
coherent reversal processes that may involve vortices \cite{Kirk01}. As a consequence, the
switching field of even very small samples may differ from the prediction of the macrospin model.
Some analytical models have been developed for soft\cite{bib-GRI01} of
hard\cite{bib-FRU01,bib-FRU04b} magnetic materials. It is moreover very likely that these effects
are amplified by the surface anisotropy term and the exchange reduction at the surfaces
\cite{Dimian02,bib-ROHtbp}.

These considerations were limited to perfect samples. The presence of some roughness, especially
at the edges of small elements patterned from thin films, was shown to have a big influence on the
switching properties\cite{Cowburn00b}. From a magnetostatic point of view, edge roughness
increases the energy of a configuration with tangential magnetization. This energy contribution
reduces the shape anisotropy, and has been called lateral interface anisotropy\cite{Cowburn00b}.
It generally reduces switching fields compared to perfect shapes.

\subsection{Confined multidomain states}

\begin{figure}[b]
  \begin{center}
  \includegraphics[width=120mm]{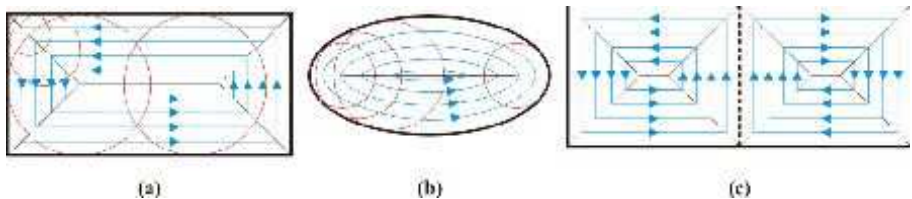}%
  \end{center}
  \caption{\label{fig-VdB}Examples of (a-b)~first order and (c)~higher order Van den Berg's constructions
    for flux-closure magnetic states.}
\end{figure}

Multidomain states, also called flux-closure-domain states, have been much studied in the
technologically relevant case of thin flat dots made of soft magnetic material like Permalloy. In
the limit of vanishing thickness, infinite lateral size and zero MAE, the shape of flux-closure
domains is predicted by the Van den Berg's~(VdB) construction\cite{bib-VAN84,bib-VAN86}, which
exhibits charge-free states~(notice that, owing to the infinite lateral size, exchange in the
domains and the energy of domain \emph{walls} are neglected): magnetic walls are located at the
locii of the centers of all disks tangent to the edge of the structure on at least two points. The
magnetization vector then lies perpendicular to their radii, \ie remains parallel to the closest
edge\bracketsubfigref{fig-VdB}{a-b}. Configurations of higher order can be obtained by dividing
the shape in several equal parts and applying the construction to each of
them\bracketsubfigref{fig-VdB}{c}. Higher order patterns can result from magnetic history, or
arise in the case of moderate in-plane anisotropy to favor domains along easy axis directions. The
VdB model was extended under in-plane field\cite{bib-BRY89,bib-BRY89b,bib-DES01}. All these
features were checked experimentally in dots (tens of) microns wide\cite{bib-HUB98b}. Surprisingly
these models also work reasonably well beyond their theoretical range of validity, \ie for
sub-micron-size, non-soft and rather thick dots, as far as the \emph{center} of the walls is
concerned\cite{bib-HER99,bib-FRU03c,bib-FRU04c}. However, a more detailed description at this
scale requires the use of micromagnetic theory. This was done analytically \eg for describing the
vortex state in disks\cite{bib-GUS01,bib-GUS01b,bib-JUB04}. Micromagnetic simulations must be used
to tackle more complex situations, like the energetics of flux-closure patterns of different
orders~(see Fig.\ref{fig-VdB}), whose degeneracy in the simple VdB's approach is lifted when the
energy of domains walls is rigourously computed\cite{bib-RAV00,bib-CHE05}. Finally, when the
thickness of a dot exceeds by far $\Delta$ and/or $\Lambda$ significant variations of
magnetization are allowed along the thickness. Such situations provide an interesting intermediate
situation between bulk materials that must be described phenomenologically, and thinner and
smaller samples that are now understood microscopically. Reports in this field are less common,
and include confined stripe domains in perpendicular media\cite{bib-HEH96,bib-HA03}, distorted VdB
patterns\cite{bib-FRU05}, 3D flux-closure in magnetically-soft cubes\cite{bib-RAV98b}.

Many experimental, simulation and theoretical reports can also be found on the hysteresis of
flux-closure mesoscopic patterns, see \eg
\cite{bib-SCH00,bib-FRU04c,bib-GUS01,bib-GUS01b,bib-WEI02,bib-RAH03}. Starting from saturation,
flux-closure domains are formed through the nucleation of vortices at the edges. Thus, like for
bulk materials, the microscopic details of nucleation remain unaccessible especially because
lithography processes often alter the edges in an ill-characterized way. Also, simulated features
may sensitively depend on the mesh used~(size, tetrahedrons in finite elements or prisms in finite
differences methods, with sometimes spurious effects on tilted edges\cite{bib-GAR03,bib-JUB04}),
the order of polynomial interpolation of the various micromagnetic quantities between the nodes of
the mesh, the minimisation algorithm used~(energy minimization or precessional dynamics with
damping). Thus, great care is needed in analyzing and comparing nucleation results, because
different nucleation events can lead through bifurcation to different flux-closure patterns at
zero field\cite{bib-FRU04c}.

One fundamental interest of confined flux-closure patterns is to benefit from the internal dipolar
field of a nanostructure that traps rigidly one or a few vortices and/or wall, and consider these
as magnetic objects. These objects can be better studied and manipulated through the application
of (possibly strong) external fields while magnetic \emph{domains} remain unaffected. In thin
films such fields often move these objects out of the field of imaging and only limited
experiments have been reported\cite{bib-HAR86,bib-THI94}. This includes the stabilization of
asymmetric N{\'e}el walls at thicknesses well beyond those found in thin films\cite{bib-FRU05}, the
topological identity of a vortex with a Bloch wall of finite length\cite{bib-HER99}, the
compression/expansion\cite{bib-WAC02} and magnetization reversal through Bloch point
nucleation\cite{bib-OKU02,bib-THI03} of vortex' cores under the application of a longitudinal
field. The following section reports on further examples of the manipulation of magnetic walls as
individual objects, in a semi-confined geometry.

\subsection{Magnetic walls in stripes}

Samples very long in one dimension but of nanometric size transversally (nanowires, nanostripes,
nanotubes) are being intensively studied, both as a challenge to
nanofabrication\cite{Piraux94,bib-NIE01,Snoeck03,Saitoh04} as for their
properties\cite{Thiaville05b} and applications. Indeed, such structures switch by the motion of
one domain wall (DW) with a well-defined velocity. Phase diagrams for the DW structures were also
computed and measured\cite{McMichael97,Nakatani05,Klaui04}. For small enough wire transverse
dimensions, it was shown that the Bloch wall model \cite{Schryer74} could be adapted to these
structures, even if they are not at all Bloch walls \cite{Thiaville05b}. One of the spectacular
consequences of these various DW structures is the predicted huge velocity difference between two
DW structures in cylindrical wires, namely the transverse and Bloch point walls \cite{Forster02}.
The dynamics of a DW under a strong current flowing along the wire, due to the spin transfer
effect, is now an active subject, both experimentally \cite{Vernier04,Yamaguchi04,Klaui05} and
theoretically \cite{Zhang05,Thiaville05}. Domain walls can also be trapped when the
cross-sectional area decreases, a so-called geometrical constriction. When this area decreases
steeply enough in the core of the constriction the wall is compressed. Its width is then predicted
to be determined solely by the geometry of the constriction, independently from materials'
parameters like exchange\cite{bib-BRU99b}. The compression in nanometer-sized constrictions has
been confirmed experimentally\cite{bib-PIE00}.

\section{Dipolar interactions}

\subsection{Coupled layers}

Two magnetic layers $F1$ and $F2$ separated by a non-magnetic layer $N$ with rough interfaces are
coupled through dipolar fields. This situation was first described by N{\'e}el, and named \emph{Orange
peel coupling}\cite{bib-NEE62}. It was later pointed out that N{\'e}el's model was developed for
semi-infinite $Fi$'s, whereas for the really thin films studied nowadays a different formula is
more adequate\cite{bib-KOO99}, predicting a much reduced coupling field $H_\mathrm{N}$. This fact
is still too often ignored. For vertically-correlated roughness the coupling is positive for
in-plane magnetization\cite{bib-KOO99}, while for perpendicular magnetization the sign of the
coupling depends on geometrical and material parameters\cite{bib-MOR04}. In all cases the coupling
decays exponentially with the thickness of the spacer layer.

Bi-(or multi-)layers of finite lateral size, \ie in the form of dots, are subject to a negative
coupling arising from the magnetic poles at the edges of the dot for in-plane magnetization, and a
positive coupling for out-of-plane magnetization. For both cases an upper bound for $H_\mathrm{N}$
arising from $F1$ or $F2$ is $N_1M_{\mathrm{s},1}$ with $N_1$ and $M_{\mathrm{s},1}$ the in-plane
demagnetizing factor and the magnetization of $F1$, respectively.

Let us examine the consequences of coupling in bilayers\cite{bib-VAN00}. Notice that the physics
described below may arise from other types of coupling, like RKKY\cite{bib-BRU99}. The limit of
weak coupling is when $H_\mathrm{N}$ is smaller than $H_\mathrm{c,1}$ or $H_\mathrm{c,2}$. In such
a case the coupling results in shifted~(biased) minor hysteresis loops. Notice however, that even
in this weak coupling limit dipolar fields may be locally much more intense than $H_\mathrm{N}$
when domain walls occur, like during magnetization reversal\cite{bib-THO00}. This may lead to
progressive demagnetization of the hard layer of spin valves\cite{bib-THO00b} or nucleation of
reversed domains in the vicinity of domain walls\cite{bib-VOG05}. In the strong coupling limit
$H_\mathrm{N}$ is larger than both coercive fields, resulting in rigidly coupled layers. The
single-domain limit is shifted upwards for in-plane magnetization in dots because demagnetizing
fields are reduced, while it is shifted downwards for out-of-plane magnetized dots. Multidomain
states are also affected as the flux might be partly closed from one layer to the other, yielding
magnetization vectors locally perpendicular to lateral edges.

\subsection{Dipole-dipole lateral interactions}

Nanostructures are often found in planar networks, see Ref.\cite{bib-MAR03} for a review. In the
point dipole approximation an upper bound for the stray field acting at a given site from
neighbors closer than radius $R$ and of arbitrary direction of magnetization is proportional to
$(\muZero/4\pi)\int_0^R \frac{2}{r^3}2\pi r \diff r\longrightarrow \mathrm{Cte} +
\mathcal{O}(1/R)$. Thus dipolar fields are \emph{short ranged} in 2D, contrary to the 3D case. To
go beyond the point-dipole approximation one can use analytical formula in the case of spheres or
prisms\cite{bib-HUB98b}, and for more complex shapes micromagnetic codes or a multipole
approach\cite{bib-MIK05}. In practice, for a regular network the range of dipolar fields scales
with the thickness of the nanostructures, which means first neighbors for \eg flat
dots\cite{bib-FRU99b}, or many neighbors \eg in the case of a dense array of elongated
cylinders\cite{bib-NIE01}. For perpendicular anisotropy dipolar fields favor
checkerboard\cite{bib-AIG98} or stripe patterns\cite{bib-NIE01} depending on the mesh symmetry,
\eg square and hexagonal, respectively. For planar magnetization alternating rows of dots with
parallel and antiparallel magnetization directions are favored along an easy axis of the
nanostructures in the presence of magnetic anisotropy, or along certain rows of the network in the
case of nanostructures magnetically isotropic in-the-plane\cite{bib-MAR98}. Even for weak
interactions such states can be approached \eg through demagnetization procedures. Formalisms and
techniques used to characterize couplings include the Preisach model\cite{bib-MAY91}, Henkel
plots\cite{bib-HEN64,bib-THA98}, or simply shifted minor loops.

\section{Thermal effects}

\subsection{Thermally activated magnetization reversal}
\label{subsec-ThermalActivation}

On time scales larger than approx.\unit[1]{ns} the effect of temperature on magnetization
processes can be fairly well described by an Arrhenius law proposed by Brown\cite{bib-BRO63} and
checked recently against LLG macrospin simulations in the range of tens of
nanoseconds\cite{bib-VOU04}: thermal energy allows to overcome an energy barrier $\Delta E$ after
a waiting time $\tau=\tau_0\exp(\Delta E/\kb T)$ with $\tau_0\approx\unit[10^{-10}]{s}$. The
non-trivial issue is to estimate $\Delta E$.

It occurs that for single-domain nanostructures not larger than the domain wall width
$\wallWidth$~(\eg nanometer-sized clusters\cite{bib-BON99} and made of soft magnetic material, the
macrospin approximation and the Stoner-Wohlfarth model roughly hold \emph{during magnetization
reversal}\cite{bib-LU99}. In the case of uniaxial anisotropy and a field applied along the easy
axis we have

\begin{equation}
  \label{eq-EnergyBarrier}
  \Delta E=KV(1-H/\Ha)^2
\end{equation}

with $\Ha=2K/\muZero\Ms$ is the anisotropy field and $V$ the volume of the nanostructure. For a
measurement performed over a time duration $\tau$ the expected coercivity is

\begin{equation}
  \label{eq-CoercivityWithTimeAndTemperature}
  \Hc(T,\tau)=\Hc(T=\tempK{0})\left({1-\sqrt{\frac{\kb T}{KV}\ln{\frac{\tau}{\tau_0}}}}\right)
\end{equation}

Such and other predictions were first confirmed experimentally using single-particle
measurements\cite{bib-WER97}. Notice that in general when $H$ is applied in an arbitrary
direction, even close to an easy axis, the dependance of $\Delta E$ with $H$ is non polynomial.
The first-order expansion of this dependence defines a generalized exponent $\alpha$: $\Delta
E=KV(1-H/\Ha)^\alpha$, with $\alpha=1.5$ in most cases \cite{bib-VIC89,Thiaville98b}.

For nanostructures larger than $\wallWidth$ \eqnref{eq-EnergyBarrier} is not valid, because
magnetization reversal if not uniform. One approach consists in replacing $V$ with a so-called
\emph{nucleation volume} $\Vn$ and consider a phenomenological generalized exponent $\alpha$.
$\Vn$ and $\alpha$ may be determined experimentally with temperature- or time-dependent
magnetization reversal, the former being sometimes ambiguous because $K$ may vary with~$T$.
Besides, time-dependent measurements may be performed at constant field~(gate functions with
variable duration) or at constant field \emph{variation}~(triangle functions). The latter
procedure is easier to implement experimentally, however the analysis is more tedious requiring
the use of models like Kurkij{\"a}rvi's\cite{bib-KUR72}, which predicts a linear variation of $\Hc$
with $\linediff{H}{t}$. Experimentally $\Vn$ is often found of size similar to
$\wallWidth$~($\wallWidth^3$ for bulk, $t\wallWidth^2$ for structures of thickness $t<\wallWidth$
\etc), and $\alpha$ is generally in the range $1-2$\cite{bib-FRU99b}. $\alpha=1$~is often found in
thin films when domain-wall propagation events determine coercivity, see Ref.\cite{bib-FER01} for
a review. Finally notice that deviations from these simple laws are observed when the dynamics are
probed over many orders of magnitude. This may arise because of a cross-over, \eg from
propagation- to nucleation-limited coercivity\cite{bib-CAM01}. It has also been proposed that in
some cases this may reveal a more complex equation than \eqnref{eq-EnergyBarrier} with a $1/H$
dependence, explained by the so-called droplet model\cite{bib-MOR05}. Generalized $(1/H)^\mu$ laws
were also reported\cite{bib-LEM98} and explained by collective effects. In all theses
phenomenological approaches the details of the inhomogeneous magnetization reversal process remain
hidden. When full micromagnetic models are available\cite{bib-FRU01,bib-BRA93,bib-BRA99} then
$\Vn$ and $\alpha$ can be evaluated directly, and it is often found that $\alpha$ results from a
fit to a non-polynomial variation, so that $\alpha$ is in fact dependent on $T$ and $\tau$.

\subsection{Superparamagnetism}
\label{subsec-Superparamagnetism}

\begin{figure}[b]
  \begin{center}
  \includegraphics[width=156mm]{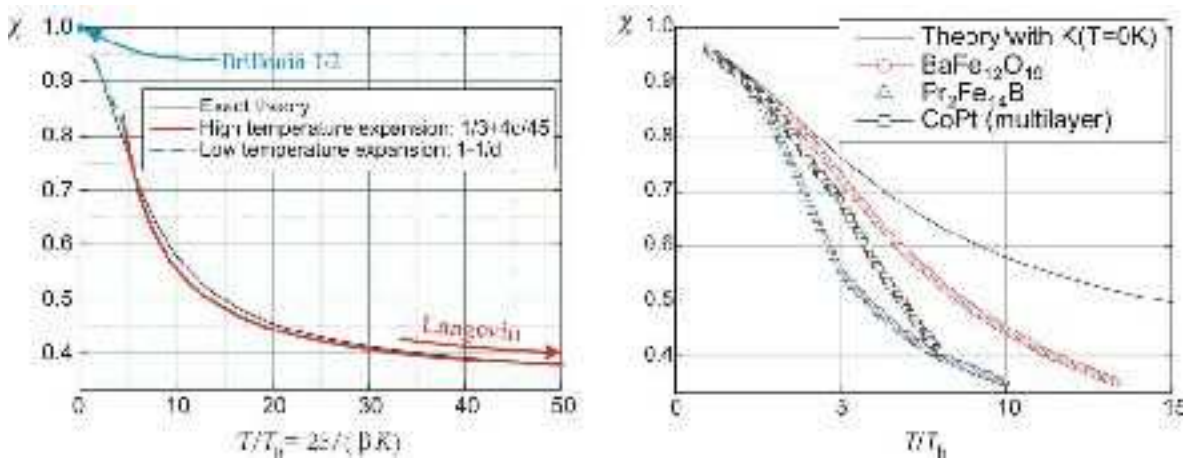}%
  \end{center}
  \caption{\label{fig-superpara}Exact result and asymptotic expansions for the initial susceptibility
  of systems with uniaxial second order anisotropy and external field applied along the easy axis direction.}
\end{figure}

\eqnref{eq-CoercivityWithTimeAndTemperature} predicts that, for a fixed time scale $\tau$, $\Hc$
vanishes for $T>\Tb$, with $\Tb=(KV/\kb)\ln(\tau/\tau_0)$ being called the \emph{blocking
temperature}. This phenomenon is called \emph{superparamagnetism}, in analogy with paramagnetism
however considering macro-(or \emph{super})spins. The variation of the time-averaged $M$ with $H$
is generally described by statistical physics using a Boltzmann occupancy law, with the magnetic
energy including Zeeman and anisotropy energies. Experimental data in the superparamagnetic regime
therefore potentially contain information on the magnetic moment and the anisotropy of the system.
However in most systems, \eg assemblies of nanoparticles, other parameters interfere like easy
axis orientation, interparticle dipolar interactions, and distributions of all these parameters.
It is then tricky, or even impossible, to perform a reliable analysis. See \cite{bib-COF93} for a
review.

Quantitative analysis is reliable only when most parameters are known. Let us concentrate on cases
with negligible interactions and the external field applied along an easy axis direction. For
vanishing anisotropy it is readily derived that the normalized average moment is
$m=\mathcal{L}(h)$ with $\mathcal{L}(h)=1/\tanh(h)-1/h$ the Langevin function with
$h=\muZero\mathcal{M}H/\kb T$ and $\mathcal{M}$ the magnetic moment of the system. However
anisotropy cannot be neglected until $T\gtrsim20\Tb$, which is seldom the case in experiments, so
that $\mathcal{L}$ should be used with care. For infinite uniaxial anisotropy the Ising case is
retrieved: $m=\mathcal{B}_{1/2}(h)$ with $\mathcal{B}_{1/2}(h)=\tanh(h)$ the Brillouin $1/2$
function. This case is relevant for self-organized systems with perpendicular anisotropy, which
received recently a considerable interest\cite{bib-FRU99d,bib-OHR01,bib-RUS03}. For the real case
of finite anisotropy the agreement with the Ising case is satisfactory for $\Tb<T\lesssim5\Tb$. An
exact expression spans all cases from infinite to vanishing anisotropy\cite{bib-CHA85,bib-FRU02b}.
Of particular interest for fitting experimental data are the first order expansions of the zero
field susceptibility in the low-temperature range ($\chi\sim 1/3+4d/45$) and high temperature
range ($\chi\sim1-1/d$) with $d=KV/\kb T$ with a cross-over around $T=5\Tb$, which match nearly
perfectly the analytical expression\cite{bib-FRU03f}\bracketfigref{fig-superpara}. Notice that we
considered temperature-independent material parameters, whereas magnetization, and even more MAE,
are expected to decays significantly with temperature in reduced dimensions. This may play a
significant or even dominant role, which was not considered here. Dipolar and other interactions
can be revealed by an offset in $1/\chi(T)$ plots\cite{bib-FRU99d}. The peak in susceptibility
measurements can also be used to determine $\Tb$\cite{bib-RUS03}.

Let us conclude with comments. First, the volume relevant for superparamagnetism is always the
total volume of the system, not a phenomenological activation volume. Second, superparamagnetism
is a drawback for applications in magnetic memory, however it is an advantage for determining
parameters of the system, provided that a relevant and robust fitting procedure is used, as
explained above. Superparamagnetism is also an advantage to prevent aggregation of ferromagnetic
nanoparticles in microfluidic or biomedicine\cite{bib-PAN03}. It has also been used in magnetic
logic schemes\cite{bib-COW00b}.

\section*{Acknowledgements}


\begin{thebibliography}{100}
\bibitem{bib-dipolarSurfaceAnisotropy}
\texttt{Notice that this pioneering approach\cite{bib-NEE54} already included
  the dipolar contribution to $\Es$, arising from the the discrete location of
  magnetic moments. This effect is covered and extended in
  Ref.\cite{bib-DRA88}}.
\bibitem{bib-XMCD-on-Dots}
\texttt{Previous pioneering studides were reported, although incomplete in
  terms of spanning of dimensionnality, see \eg
  \cite{bib-DUR99,bib-KOI01,bib-OHR01,bib-GAM02b}}.
\end{thebibliography}

\end{document}